# Evidence for Flat Band Dirac Superconductor Originating from Quantum Geometry

Haidong Tian[1], Shi Che[1], Tianyi Xu[2], Patrick Cheung[2], Kenji Watanabe[3], Takashi Taniguchi[4], Mohit Randeria[1], Fan Zhang[2], Chun Ning Lau[1*], Marc W. Bockrath[1*]

[1] Department of Physics, The Ohio State University, Columbus, OH 43221.
[2] Department of Physics, The University of Texas at Dallas, 800 West Campbell Road, Richardson, Texas 75080-3021, USA
[3] Research Center for Functional Materials, National Institute for Materials Science, 1-1 Namiki, Tsukuba 305-0044, Japan
[4] International Center for Materials Nanoarchitectonics, National Institute for Materials Science, 1-1 Namiki, Tsukuba 305-0044, Japan

**In a flat band superconductor, the charge carriers' group velocity $v_F$ is extremely slow, quenching their kinetic energy. The emergence of superconductivity thus appears paradoxical, as conventional BCS theory implies a vanishing coherence length, superfluid stiffness, and critical current. Here, using twisted bilayer graphene (tBLG) [1,2], we explore the profound effect of vanishingly small $v_F$ in a Dirac superconducting flat band system[3-7]. Using Schwinger-limited non-linear transport studies[8,9], we demonstrate an extremely slow $v_F \sim 1000$ m/s for filling fraction $\nu$ between -1/2 and -3/4 of the moiré superlattice. In the superconducting state, the same velocity limit constitutes a new limiting mechanism for the critical current, analogous to a relativistic superfluid[10]. Importantly, our measurement of superfluid stiffness, which controls the superconductor's electrodynamic response, shows that it is not dominated by the kinetic energy, but instead by the interaction-driven superconducting gap, consistent with recent theories on a quantum geometric contribution[3-7]. We find evidence for small pairs, characteristic of the BCS to Bose-Einstein condensation (BEC) crossover[11-13], with an unprecedented ratio of the superconducting transition temperature to the Fermi temperature exceeding unity, and discuss how this arises for very strong coupling superconductivity in ultra-flat Dirac bands.**

The dominance of electronic interactions in a flat band electronic system leads to correlated phenomena such as Mott insulators[1], Wigner crystallization[14-18], and magnetism[19-22]. Superconductivity therein is particularly intriguing, being related to the long-standing mysteries of high temperature superconductors[23] and heavy fermion systems[24]. Consequences of vanishing group velocity in a Dirac superconductor are starting to be appreciated theoretically[3-7], but yet to be observed in any experiment.

---

[*] Email: lau.232@osu.edu; bockrath.31@osu.edu

The amazing discovery of superconductivity[1,2] in tBLG has ushered in a new class of moiré-induced flat band superconducting systems[25,26]. Our tBLG devices are fabricated by the "cut-and-stack" technique on Si/SiO$_2$ substrates[27-31]. Figure 1a displays the longitudinal resistance $R$ at zero bias as a function of back gate voltages $V_g$ (bottom axis) and filling fraction $\nu$ (top axis) at $T$=0.3K. Sharp peaks emerge at carrier density $n$=0, -1.4, -2.2 and -2.8 x 10$^{12}$ cm$^{-2}$, corresponding to $\nu$=0, -1/2, -3/4 and -1, respectively, in agreement with previous works[19,26,32]. The device's twist angle $\theta$ is estimated to be 1.08°. All data are taken at 0.3 K unless otherwise specified.

We first examine the device's non-linear transport properties in the normal state, ensured by applying a small perpendicular magnetic field $B$=0.2 T. Figure 1b-c plots $dV/dI$ vs. $n$ and bias current density $J$, displaying a "bell"-like feature. The signal is low and constant at small bias ($dV/dI \lesssim 200$ Ω, blue region,), but increases dramatically to a pronounced peak at certain critical current $J_{cn}$ (yellow "lips" that line the blue region) before settling to a high resistance state ($dV/dI$ ~ 5-7 kΩ, green region). For -1/2<$\nu$<-5/8, $J_{cn}$ increases almost linearly with increasing hole doping. Similar critical current-like behavior in $dV/dI$ has recently been observed in tBLG with $\theta$=1.23°, in graphene/BN superlattices, and in graphene constrictions[8]. They arise from the Schwinger mechanism, also known as Zener-Klein tunneling[9,33,34] – even when the charge carriers' normal state drift velocity $v_n$ reaches $v_F$, a large electric field can create electron-hole pairs by driving charges from the valence band to the conduction band; this is facilitated by graphene's gapless Dirac spectrum, and gives rise to peaks in $dV/dI$ when $v_n \approx v_F$.[35]

Here we take advantage of the Schwinger-induced features to measure $v_F$ experimentally. Strikingly, $J_{cn}$ ~70-500 nA/μm in our device, more than two orders of magnitude smaller than that in previous reports[8]; this reduction is fully expected since our device is at the magic angle. From one-sided Landau fans and resetting of the Hall resistance at $\nu$=-1/2, we find that the Dirac band resets at this half filling, in agreement with prior works[19,26,32]. Hence, in the following the effective charge density $\tilde{n}$ is measured from $\nu$=-1/2. Using $v_F = J_{cn}/\tilde{n}e$, we obtain $v_F$ values ranging from ~700 to 1200 m/s (Fig. 1d). This extremely slow velocity is reduced from the Fermi velocity of monolayer graphene 10$^6$ m/s by three orders of magnitude, indicating that the Fermi level of the superlattice is ~1 meV, constituting the flattest minibands reported to date. The charge carriers' effective mass, estimated as $m=\hbar k_F/v_F$, is exceedingly heavy, reaching~30 $m_e$ at $\tilde{n}$=-5x10$^{11}$ cm$^{-2}$ (here $\hbar$ is the Planck's constant, $k_F = \sqrt{2\pi\tilde{n}}$ the Fermi wave vector, where the factor of 2 arises from the 2-fold quasiparticle band degeneracy near the half-filling point[1,19,26,32], and $m_e$ the rest mass of electron) (Fig. 1d). Our band structure calculations[35] verify that $v_F$~1000 m/s is reached at mini-Dirac points in devices with $\theta$=1.08°, close to the experimental value (Fig. 1e-f).

The ultraflat band with a miniscule bandwidth is further confirmed by temperature $T$-dependent measurements. When $k_BT$ becomes comparable to the bandwidth, additional electron-hole pairs are thermally excited, and the transition from regular to Schwinger-dominated transport becomes smeared, suppressing the $dV/dI$ peaks[8]. Indeed, as $T$ increases from 0.3K, $J_{cn}$ remains constant, but the $dV/dI$ peaks are smeared and suppressed (Fig. 1g-h); at $T$~6K, $dV/dI$ features are barely distinguishable[35], in agreement with an ultra-narrow bandwidth of <~1 meV.

After establishing the ultra-flat band and low $v_F$ in the device, we now focus on transport data at $B=0$. Superconductivity is observed at $-3.5<\tilde{n}<0.3 \times 10^{11}$ cm$^{-2}$, with a characteristic dome shape in the $T$-$\tilde{n}$ plane (Fig. 2a-b). Here we define the superconducting Berezinskii-Kosterlitz-Thouless (BKT) transition temperature $T_c$ [36,37] to be the temperature at which $R$ first exceeds 20% of the normal state resistance at zero bias. $T_c$ reaches ~2.2K at the top of the dome, $\tilde{n}$~ -1.8 x 10$^{11}$ cm$^{-2}$, taken to be the optimal doping point. Following the convention of cuprates, we refer to the regions with charge density to the right and left of the optimal doping as underdoped and overdoped, respectively.

Similarly, a superconducting dome is observed vs. $B$ and $\tilde{n}$ (Fig. 2c), where the critical magnetic field $B_{c2}$ reaches as high as ~0.1T at optimal doping (Fig. 2d, left axis). Taking this critical field value as the upper critical field, we estimate that the superconducting coherence length $\xi = \sqrt{\frac{\Phi_0}{2\pi B_{c2}}}$ to be ~55 nm at optimal doping, where $\Phi_0$ is the flux quantum, and increases to hundreds of nm when $\tilde{n}$ approaches half-filling or the von Hove singularity (Fig. 2d, right axis). These values are consistent with prior reports[2,19,26,29,38].

From these $\xi$ and $v_F$ measurements, several fundamental BCS relations are clearly invalidated. For example, a fundamental energy scale for the pairing gap is given by $\Delta \sim \hbar v_F / \xi$. Using $v_F$~1000 m/s and $\xi$=55 nm at optimal doping, $\Delta/k_B T_c$ is found to be ~0.05, far smaller than the ratio of 1.75 in conventional superconductors. Similarly, Pippard's argument using the uncertainty principle to obtain $\xi \sim \Delta x \sim \frac{\hbar v_F}{k_B T_c}$, yields $\xi$~ 2.6 nm at optimal doping in this device, much smaller than that measured from $B_{c2}$ data. Another fundamentally important quantity is the superfluid stiffness $D_s$ that determines the superconductor's electromagnetic response. Conventionally, $D_s(T)=e^2 n_s(T)/m$. Assuming all the electrons contribute to the superfluid density, $n_s(0)=\tilde{n}$, we find $D_s(0)$ ~10$^6$ H$^{-1}$ at optimal doping. This yields an upper bound of BKT transition, as $D_s$ controls the loss of phase coherence through the Nelson-Kosterlitz criterion[39]

$$\frac{\hbar^2 D_s(0)}{e^2 k_B T_c} \geq \frac{8}{\pi} \qquad (1)$$

Our estimated $D_s(0)$ thus yields $T_c \leq 0.01$ K at optimal doping, far below the observed $T_c = 2.2$ K. All these invalidated equations therefore indicate that the flat-band superconductivity is markedly different from the conventional (BCS-like) behavior. We will return to this point in subsequent discussion.

We now focus on non-linear transport in the superconducting state. Fig. 3a plots $dV/dI$ vs. $I$ and $\tilde{n}$. For $|\tilde{n}|>4.0 \times 10^{11}$ cm$^{-2}$, the device is metallic, and the $dV/dI$ peaks are identical to those at $B=0.2$T (Fig. 3b). The most interesting features, however, occur in the superconducting state. Surprisingly, the same "bell-like" features persist, with almost identical outlines of critical current density as that in Fig. 1b. However, two important differences between $dI/dV$ data at $B=0$ and $B=0.2$T exist. First, throughout the underdoped region, the high-bias peaks are extremely sharp; under a small magnetic field, the peak amplitudes are suppressed and recover their normal state values. We therefore identify the current density corresponding to these very sharp peaks as the superconducting critical current density $J_{cs}$. Importantly, $J_{cs} \approx J_{cn}$ for this underdoped region, i.e.

*the peak positions are almost identical in both superconducting and normal phases* (Fig. 3c-d). Second, at higher doping ($\tilde{n} < -2.0 \times 10^{11}$ cm$^{-2}$), the *dV/dI* peak bifurcates –the outer peak has similar location and amplitude as that in the normal state, whereas the inner peak occurs at a smaller current density. A small magnetic field suppresses the inner peak, thus we identify the corresponding current density to be $J_{cs}$ (Fig. 3e); $J_{cs}$ decreases rapidly with increasing doping in this overdoped region, and vanishes at $\tilde{n} \sim -3.8 \times 10^{11}$ cm$^{-2}$.

Figure 3f plots $J_{cs}$ and $J_{cn}$ as red circles and blue triangles, respectively. The bifurcation of $J_{cn}$ and $J_{cs}$ in the overdoped regime suggest two distinct mechanisms limit the superconducting critical current. However, their coincidence in the underdoped regime indicates that the same current-limiting mechanism in the normal state, namely the band velocity limit in a Dirac system, limits the current density in the superconducting phase.

We first examine the limit to $J_{cs}$ arising from the depairing condition[40] – a supercurrent with uniform velocity $v_s$ shifts the energy of quasi-particle excitations by $\hbar k_F v_s$, and superconductivity is destroyed when this energy shift exceeds $\Delta$, leading to a critical current density

$$J_{cs} = n_s e \frac{\Delta}{\hbar k_F} = n_s e \frac{\alpha k_B T_c}{\hbar k_F} \qquad (2)$$

where $n_s \sim \tilde{n}$ at low temperatures, and $\Delta = \alpha k_B T_c$. Assuming $\alpha \sim 2$ across the entire doping range[2,41,42], we find that the depairing critical current obtained from using Eq. (2) (Fig. 3f, dotted black curve) is at least an order of magnitude higher than the measured $J_{cs}$. We emphasize that this depairing value is the lower limit, since $\Delta / k_B T_c$ is typically higher in superconductors that are unconventional, or in the strong coupling limit[11,12].

Now we consider the new limiting mechanism to the supercurrent, i.e. the vanishingly small $v_F$. The depairing condition in a conventional superconductor implicitly assumes that (i) the energy shift $\hbar k_F v_s$ represents a small perturbation of the Fermi sea, and (ii) the band dispersion is quadratic, so that there is no saturation limit to $v_s$. However, neither of these assumptions is valid in tBLG with extremely flat Dirac minibands. In a gapless Dirac spectrum, while the condensate's momentum and velocity are proportional at small momentum, the velocity saturates to the band velocity $v_F$ at large momentum, as illustrated in Fig. 3g. As the current density increases, the phase gradient of the order parameter continuously grows, while $v_s$ asymptotically approaches $v_F$, as depicted in Fig. 3i-j. This is analogous to the acceleration of a particle in special relativity, where the relativistic mass continuously increases while its speed asymptotically approaches the speed of light. Once $\nabla \varphi \sim 1/\xi$, where $\xi$ is the coherence length, superconductivity is destroyed, and $v_s \approx v_F$. Beyond this point, electron-hole production via the Schwinger mechanism allows additional current to flow.

In the underdoped regime, such velocity saturation is expected to be reached prior to the depairing limit; in the overdoped regime, however, as $\Delta$ diminishes, we recover the depairing condition Eq. (2) as the primary limiting mechanism. A full theoretical treatment of the critical current is beyond the scope of the work. Nevertheless, we can account for the fact that $J_{cs}$ is determined by the *smaller* of these two limits by phenomenologically writing the effective velocity as

$$v^{-1} = v_F^{-1} + \left(\frac{\Delta}{\hbar k_F}\right)^{-1} \qquad (3)$$

and calculate $J_{cs}=\tilde{n}ev$. The green dashed line in Fig. 3f shows the resulting curve, in reasonable agreement with the experimental data, thus confirming the presence of both conventional and unconventional limits to supercurrent density in magic angle tBLG.

In addition to demonstrating a new limiting mechanism to the supercurrent, the observed $J_{cs}$ also permits us to estimate the superfluid stiffness $D_s$ in this flat-band superconductor. As discussed above, the conventional estimate of $D_s(0)$ yields $T_c \leq 0.01$ K at optimal doping, far below the observed $T_c = 2.2$ K; it also increases monotonically with $\tilde{n}$ (Fig. 4a, black dotted line), instead of displaying the dome-like shape. Clearly, the actual $D_s$ must be far larger than that based on this naïve estimate.

Since conventional techniques of measuring $D_s$ cannot be applied to mesoscopic devices, we extract $D_s$ from our data by starting from the basic equations that relate $D_s$ and the gauge invariant momentum, $\vec{p} = \hbar \nabla \varphi - 2e\vec{A}$, to the supercurrent $\vec{J} = \frac{D_s}{2e}\vec{p}$, where $\varphi$ is the phase of the superconducting order parameter and $\vec{A}$ is the vector potential. The vortex core radius $\sim \xi$ is determined by the condition where the circulating current density reaches its critical value $J_c$[43] (Fig. 4a inset). Using $|\nabla \varphi| \approx 1/\xi$ we get $J_{cs} \approx \frac{D_s \Phi_0}{2\pi\xi}$, where $\Phi_0 = h/2e$ is the flux quantum. The measured $J_{cs}$ and $\xi$ then allow us to estimate

$$D_s(0) = \frac{2\pi J_{cs}\xi}{\Phi_0} \qquad (4)$$

The extracted $D_s(0)$ follows a dome-like behavior with a maximum $\sim 5 \times 10^7$ H$^{-1}$ (solid red line in Fig. 4a). Putting this value into the $T_c$ bound in Eq. (1) yields $T_c \sim 0.6$ K without free parameters. This value is lower than the measured value of $T_c = 2.2$ K at optimal doping; nevertheless, they agree within the same order of magnitude, particularly considering the simplicity of our model. Our data provide strong evidence that the superfluid stiffness is enhanced beyond the conventional expectation based on band dispersion alone.

To understand the measured $D_s$ we turn to recent insights from mean field theories[3-7] and exact bounds[7,44] for the superfluid stiffness in flat band systems. Here the conventional contribution to $D_s$ arising from band dispersion is absent, consistent with the negligibly small experimental estimate presented above. The diamagnetic response is then determined by the quantum geometry of the flat band wavefunctions through the trace of the quantum metric or the non-trivial band topology for tBLG[45-48]. The scale of $D_s$ is set not by the kinetic energy but rather by the interactions, thus it scales with $\Delta$ [3-7,49]. Absent a theory for the full density dependence of superconductivity in tBLG, including resets at correlated insulator states, we use dimensional analysis to estimate $D_s(0, \tilde{n}) \approx b\frac{e^2}{\hbar^2}\Delta(0, \tilde{n})$, where $b$ is a constant of order unity. Using $\Delta(0, \tilde{n}) \approx 2T_c(\tilde{n})$ and $b=0.33$, we obtain the dotted curve Fig. 4a, which has the same general dome-shaped curve as that extracted experimentally. This qualitative agreement provides strong evidence that the superfluid stiffness is dominated by the interaction-driven quantum geometric contribution in tBLG, rather than the conventional contribution with a scale set by band dispersion.

Superconductivity in an ultra-flat band system, where interactions are comparable to or exceed the bandwidth, is expected to be very strongly coupled. Examining the ratio of pair size, estimated by the coherence length $\xi$, to the inter-particle distance $1/k_F$ (Fig. 4b), we find very small values of $k_F\xi$ that between 1 (underdoped) to 10 (overdoped), characteristic of the strong coupling regime of the BCS to BEC (Bose-Einstein condensation) crossover[11-13]. This is also consistent with recent observations of a pseudogap in scanning tunneling microscopy studies of tBLG[50,51].

We next plot the ratio of the superconducting $T_c$ to the Fermi temperature $T_F$ plotted vs $\tilde{n}$ in Fig. 4c. Remarkably, $T_c/T_F$ exceeds 1 for almost the entire dome (except for the very edge of the overdoped regime) and in the underdoped regime $T_c/T_F \gg 1$. The large $T_c/T_F$ arises from the very small density $\tilde{n}$, due to the reset at half filling, combined with the extremely small $v_F$, so that $T_F \sim 0.24$ K even at optimal doping. Such large values of $T_c/T_F$ are unprecedented and completely different from all other superconductors in the Uemura plot[2,52] of log $T_c$ vs. log $T_F$. Although $T_F$ may be hard to define unambiguously in strongly correlated multiband systems, in tBLG there is a natural definition in terms of the density $\tilde{n}$ measured from the reset Dirac cone at half filling. Note, however, that this $T_F$ is not the "bare" value related to the total density in the eight low-energy bands of tBLG, but already renormalized by the interactions that lead to the reset.

We can gain insight into the large $T_c/T_F$ ratio from the exact upper bounds[7,44] on the 2D $T_c$. First, as emphasized in ref. [44], $T_F/8$ is an upper bound for $T_c$ *only* for parabolic dispersion, while the general bound on $T_c$ is in terms of the optical sum rule. In a gapless Dirac system (even if $v_F$ is not small), there is finite optical spectral weight from inter-band transitions in the limit of vanishing $T_F$, i.e., when $E_F$ coincides with the Dirac nodes. Thus, a superconducting Dirac system does *not* have its $T_c$ limited by $T_F$. Second, $T_F$ is vanishingly small for any flat band superconductor (Dirac or otherwise), but in a superconducting Dirac system the quantum geometric contributions to the low-energy optical spectral weight and $D_s$ are finite[7,44] and lead to a finite $T_c$. We thus understand why tBLG, a flat band Dirac system, is in an unprecedented regime with $T_c/T_F > 1$.

In conclusion, we have shown that tBLG is a highly unusual superconductor whose properties cannot be understood within conventional BCS theory. The Schwinger mechanism limited non-linear transport in the normal state gives evidence for a Dirac band with extremely slow Fermi velocity $v_F$. Using this $v_F$ and charge density measured from the reset, we show that the superfluid stiffness is dominated by quantum geometric contributions, the coherence length is of order the interparticle separation of electrons, and the transition temperature exceeds the Fermi temperature over much of the superconducting dome. Our experimental results uncover the mysteries of very strongly coupled superconductivity in an ultra-flat band Dirac system. The ability to tune density and temperature may allow an understanding of this novel phase of matter with connections to many areas of physics including the BCS-BEC crossover in relativistic systems, neutron stars, and quantum chromodynamics.

**Acknowledgement**
We thank Petr Stepanov for advice on device fabrication. The experiments are supported by DOE BES Division under grant no. DE-SC0020187. M.R. and the nanofabrication facility were supported by NSF Materials Research Science and Engineering Center Grant DMR-2011876. T.X., P.C., and F.Z. were supported by the Army Research Office under grant number W911NF-18-1-0416 and the National Science Foundation under grant numbers DMR-1945351 through the CAREER program. Growth of hexagonal boron nitride crystals was supported by the Elemental Strategy Initiative conducted by the MEXT, Japan (Grant Number JPMXP0112101001) and JSPS KAKENHI (Grant Numbers 19H05790, 20H00354 and 21H05233).

**Figure 1. Normal state transport of a tBLG with $\theta=1.10°$ at $B=0.2$T and $T=0.3$K (unless specified otherwise). a.** Log plot of longitudinal resistance $R$ vs. $V_{bg}$ (bottom axis) and filling fraction $\nu$ (top axis) at $B=0$. The shaded region is the range of density that we focus on. **b-c.** $dV/dI$ $(J, \tilde{n})$ in k$\Omega$, and line traces at different $\tilde{n}$. **d.** Extracted $v_F$ (red curve, left axis) and effective mass (blue curve, right axis) vs. $\tilde{n}$. **e-f.** Computed band structure of a tBLG with $\theta=1.08°$ and $v_F\sim1000$ m/s near the mini-Dirac points. **g-h.** $dV/dI(J,T)$ at $\tilde{n}=-1.75\times10^{11}$ cm$^{-2}$, showing the smearing of the peaks with temperature. Line traces are taken at $T=0.34$K, 2K, 3K, 4K, 5K and 5.8K, respectively (blue to red).

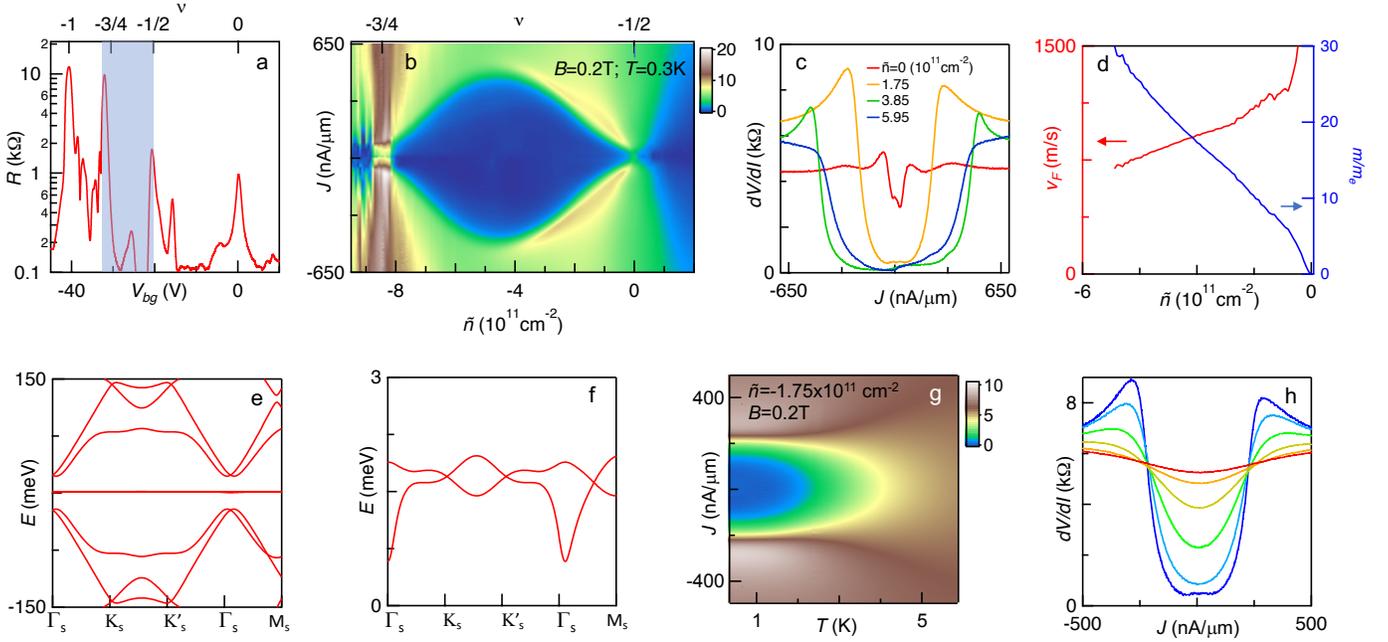

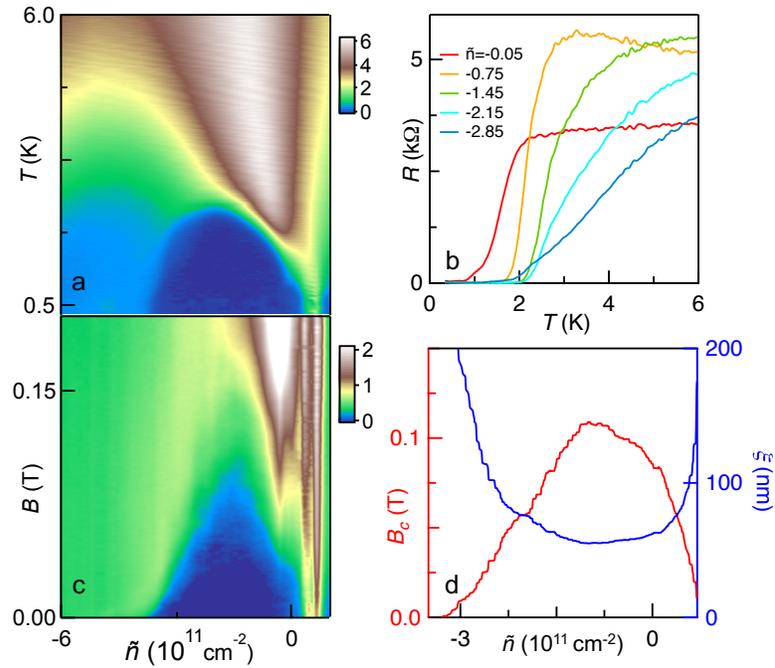

**Figure 2. Zero-bias transport data at $B=0$. a-b.** $R(T,\tilde{n})$ in k$\Omega$ showing the superconducting dome (dark blue), and line traces $R(T)$ at different $\tilde{n}$. **c.** $R(B, \tilde{n})$ at $T=0.3$ K. **d.** Critical magnetic field $B_c$ and superconducting coherence length $\xi$ vs. $\tilde{n}$.

**Figure 3. Non-linear transport data in the superconducting regime. a.** $dV/dI$ ($J$, $ñ$) in kΩ. The dotted line outlines the superconducting region for $J>0$. **b-e.** $dV/dI(J)$ at $B=0$ and $B=0.2$T, and different densities ($ñ$ labelled in units of $10^{11}$ cm$^{-2}$), respectively. **f.** Extracted critical current density in superconducting (blue) and normal (red circles) states. The black dotted line is calculated using the depairing condition Eq. (2), and green dotted line using (3) by taking both depairing and velocity saturation into account. **g.** Schematic of Fermi energy shift in a conventional superconductor with a quadratic dispersion and $\Delta<<E_F$. **h-j.** Schematics of Fermi energy in a Dirac band with small $v_F$ in the limits of $J=0$, small $J$ and large $J$ near velocity saturation, respectively. In **g-i**, charges condense below the superconducting gap $\Delta$. For simplicity of illustration, $\Delta$ and thermal smearing of the charge distribution are not shown.

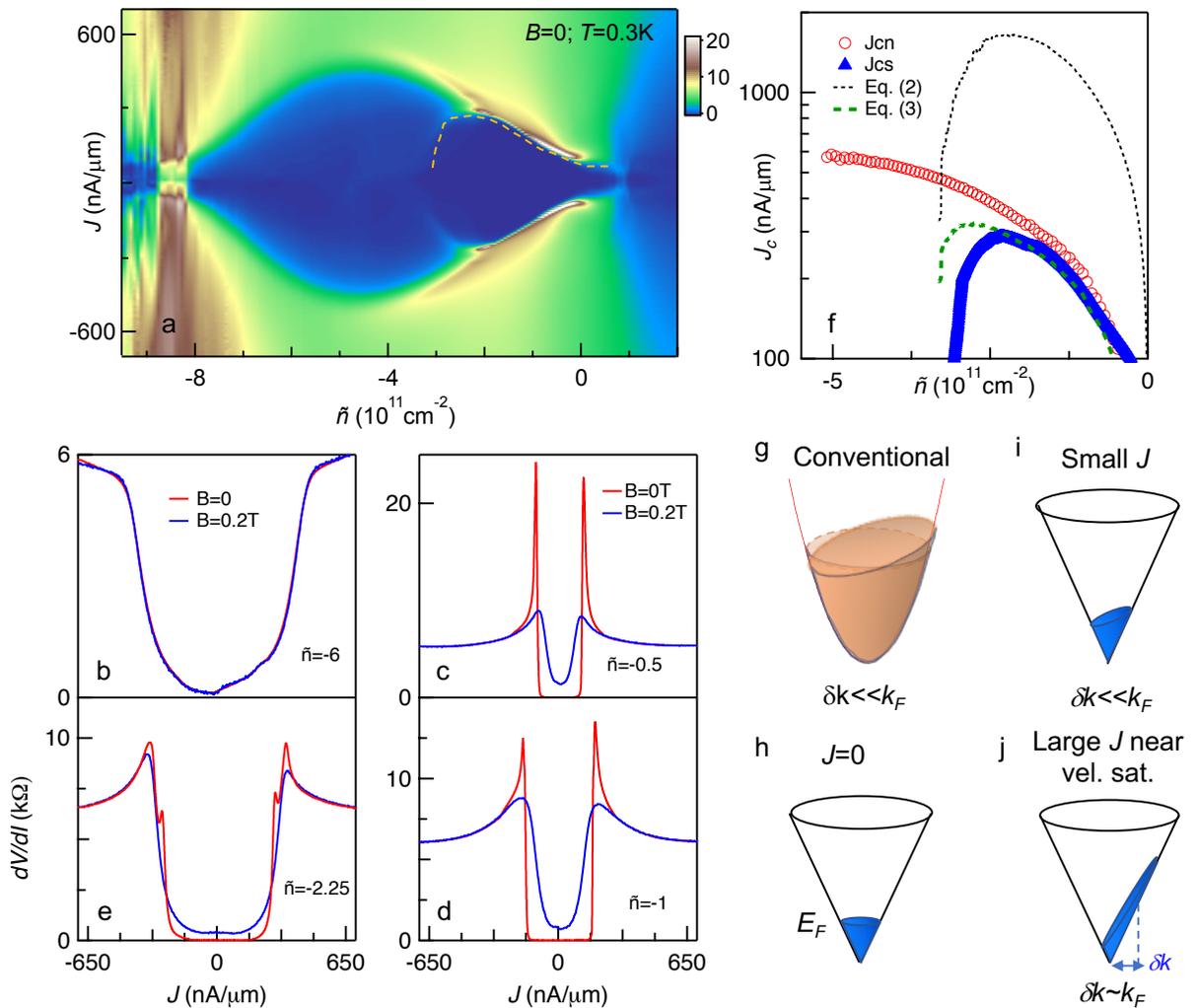

**Figure 4. Superfluid stiffness and characteristic temperatures of the flat band. a.** Inset: Supercurrent density circulating around a vortex core with radius $\xi$. Main panel: Superfluid stiffness $D_s$ vs. $\tilde{n}$. The red line is calculated from Eq. (4) using experimentally measured $J_c$ and $\xi$. The black dotted line is calculated using the conventional expression $D_s(T)=e^2 n_s(T)/m$, and the green dotted line using $D_s(0,\tilde{n}) \approx b \frac{e^2}{\hbar^2}\Delta(0,\tilde{n})$ with $b=0.33$. **b.** $k_F\xi$ vs. $\tilde{n}$ at $T=0.3$ K. **c.** $T_c$ and $T_F$ (right axes) and their ratio (left axis) vs. $\tilde{n}$.

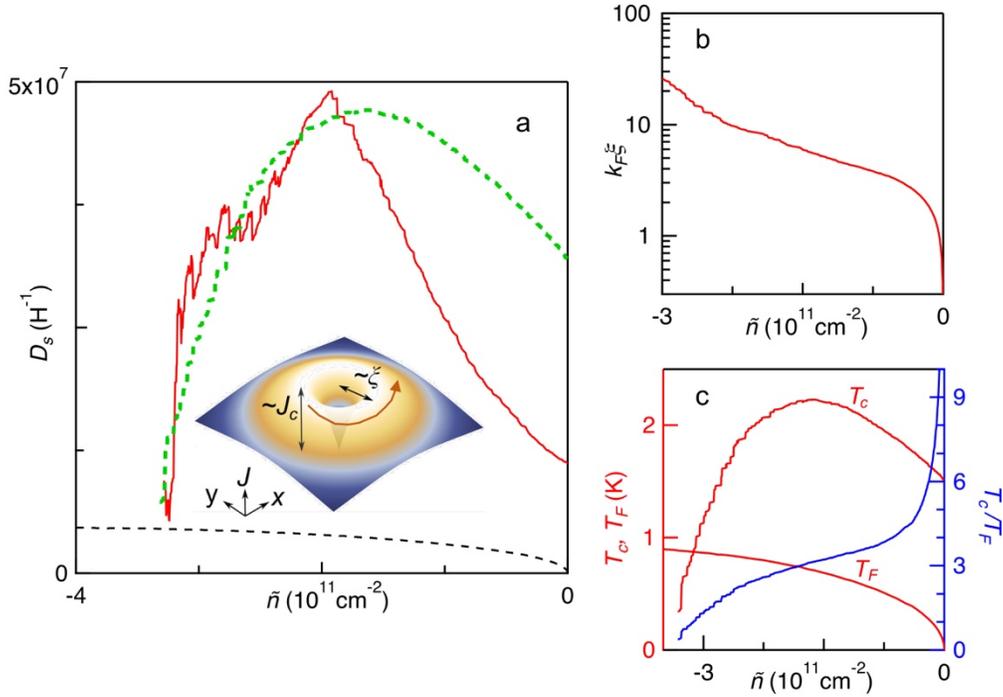